\documentclass{ws-p8-50x6-00}

\def\f{\phi}
\def\x{r/R_\odot}
\def\lesssim{\mathrel{\vcenter{\offinterlineskip\halign{\hfil
$\displaystyle##$\hfill\cr<\cr\sim\cr}}}}

\def\ARAA{\em Ann. Rev. Astron. Astrophys.}
\def\RMP{\em Rev. Mod. Phys.}
\def\ApJ{\em Astrophys. J.}

\begin{document}

\title{The Solar Core and Solar Neutrinos}

\author{Dallas C. Kennedy}

\address{Department of Physics, University of Florida, Gainesville FL 32611,
USA\\E-mail: kennedy@phys.ufl.edu $\bf{\ \bullet\bullet}$ WWW: http://www.phys.ufl.edu/~kennedy}


\maketitle

\abstracts{The long-standing deficit of measured versus predicted solar neutrino fluxes
is re-examined in light of possible astrophysical solutions.  In the last decade, solar 
neutrino flux and helioseismic measurements have greatly strengthened the case for 
non-astrophysical solutions.  But some model-independent tests remain open.}

The solar neutrino problem has nagged physicists for over 30 years and, for most of us, has a
natural solution in neutrino oscillations, either {\em in vacuo} or matter-enhanced (MSW 
effect).  This modification of neutrino properties (with $m_\nu\lesssim 10^{-2}$ eV and
mixing $\sim$ 0.001--1) requires an extension of the Standard Model not detectable in 
accelerators and consistent with many models of unification.~\cite{BHKL93}  But confidence in such a 
solution is based on a prior exclusion of astrophysical or nuclear physics explanations; 
only in the past decade has such an outcome become strongly credible.

Predictions of the solar neutrino flux $\f$ are outputs of a solar model, which in turn are
special cases of hydrogen-burning (4$\cdot$H $\rightarrow$ ${}^4$He) main sequence stellar models
(almost all $pp$ chain, with small CNO contribution).  These predictions are usually quoted from a particular 
model (here the Bahcall-Pinsonneault 1998 or BP98 model), and detailed models agree if the same inputs are
used.~\cite{BP98,TC98}  But a more generic
approach is attractive if it can free us from a specific model with fixed
parameters.  Greater generality is all the more important if it reveals basic properties of
the Sun and $\f$ that depend only on simple properties.
Here I outline the results of such an approach.~\cite{BK96,BK99,K00}  
As far as observations now go, it confirms the results of detailed solar models but also
specifies crucial solar observations that remain to be filled in.

\begin{figure}[t]
\begin{center}
\epsfxsize=20pc 
\epsfbox{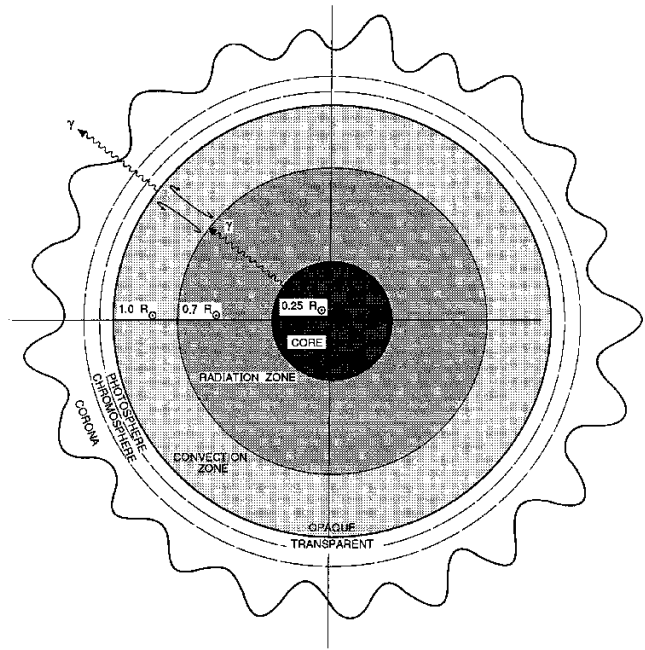} 
\end{center}
\caption{Schematic cutaway of the Sun in a generic standard solar model.~\protect\cite{KW,BU88}
\label{fig:cutaway}}
\end{figure}

\section{Properties of Solar and Stellar Models}

Simple solar models orient us with gross structure: the core $(\x\lesssim$ 0.3, 
where nuclear fusion generates the luminosity $L_\odot )$, the radiative zone (0.3 
$\lesssim\x <$ 0.71) and the convective zone (CZ, $\x >$ 0.71) (Figure~\ref{fig:cutaway}).  
Less massive stars $(M < M_\odot)$ have, according to stellar models, even deeper convective zones; while
for $M > M_\odot$, the outer convective zone disappears, and a convective inner core appears as
the central temperature gradient surpasses a critical value.~\cite{KW} 
The Sun might have had a convective inner core, 
drastically changing the $\f$ predictions, but there is now decisive evidence against core convection 
(see Section~\ref{sec:obs} below).  A higher central temperature would also have led to CNO-dominance
and {\em higher} neutrino fluxes.~\cite{BU88}

A solar model is {\em standard} (an SSM) if the model contains all the physics of matter, gravitation,
and nuclear fusion needed to obtain a star, but nothing more.  Conceptually,
stellar structure and evolution divide into three levels by time scales.  
For the Sun, chemical evolution needs $\sim$ 10 Gyr; thermal equilibrium, about
10 Myr (Kelvin-Helmholtz time); and hydrostatic equilibrium, about 5 minutes.
The hydrodynamic time controls the helioseismic $p$- and $g$-modes.  (Late in the Sun's
evolution, the chemical time scale will be shortened and the hierarchy blurred.)
{\em Structure} means only the thermal and mechanical features of a star.~\cite{KW,Arnett}

The properties of matter needed are {\em constitutive relations} giving pressure $P$, opacity
$\kappa$, and specific luminosity generation $\varepsilon$ as functions of density $\rho$ and
temperature $T$.  If the thermal structure is given, then the equation of state reduces to a 
{\em barytrope} $P$ = $P(\rho )$, and the mechanical structure alone becomes a  closed problem 
characterized by the {\em stiffness} profile $\Gamma$ = $d\ln P/d\ln\rho$.  The special case
of $\Gamma$ = $\gamma\equiv 1+1/n$ = constant is a {\em polytrope} of index $n$.
All $n <$ 5 polytropes have finite mass and radius.  $n$ = 0 is the constant-density case.~\cite{BK99,KW}

The initial conditions are fixed mass $M$ and the
element mass abundances $X_i$ (the zero-age star assumed chemically homogeneous).  The boundary
conditions are zero mass and luminosity at the center and (nearly) zero pressure and density at the 
surface.  The $X_i$ develop gradients by evolution, as nuclear fusion and heavy element diffusion act,
and additional helium accumulates in the core.~\cite{BP98,KW}

Even within the SSM framework, variations are possible.  The $\kappa$ and $\varepsilon$ functions and
fusion cross sections must be calculated from atomic and nuclear physics and extrapolated into regimes 
not directly testable.  For the dominant luminosity-producing $ppI$ reactions (terminating through 
${}^3$He--${}^3$He fusion to ${}^4$He), $\varepsilon$ and the reaction rates are almost fixed by $L_\odot$.
But reactions not strongly connected with $L_\odot$ are not well-constrained: the $ppII$ and $ppIII$
chains (terminating from ${}^7$Be through ${}^7$Li and ${}^8$B to $2\cdot{}^4$He, respectively) are
sensitive functions of the core temperature and nuclear cross sections, as well as mildly dependent 
on the core density.  $L_\odot$ and other global properties place only weak constraints on the $ppII$
and $ppIII$ rates.~\cite{BK96,KW,BU88}

\section{Generalizing the Standard Solar Model}

Generic properties of solar structure are restricted by boundary conditions.  The SSM can then
be generalized in two ways.  One is to calibrate with the specific model.  This procedure 
defines a generalized SSM family, although not the most general.

Its most convenient implementation is through {\em homology} or
power-law scaling, derivable analytically or made evident by numerical solutions.  Exploration of
model space by varying SSM inputs is actually a special case of homology, which amounts to ``small''
perturbations of the {\em logarithms} of inputs and outputs.  Such ``perturbative'' analysis works 
over a surprisingly large range, so long as the power law-relations are stable.~\cite{BK96,KW}

The first signs of homological behavior in SSMs were found in the 1000 SSM Monte Carlo study of 
Bahcall and Ulrich.~\cite{BU88}  Subsequent work over a wider model range revealed a much broader 
validity for homology.  The underlying analytic structure was derived by Bludman and Kennedy.~\cite{BK96}
Starting with structure alone, one keeps only dimensional and scaling behavior of macroscopic variables,
dropping the differential nature of the equations.  Assuming multifactor power laws for the equation
of state, opacity, and luminosity generation, we found homological relations for the mechanical and thermal 
structure, assuming fixed powers.  This requirement restricts the homology to the radiative and 
core regions, below the CZ.  The dominant luminosity production is by $ppI$, carried outwards
entirely by radiative diffusion.  The constitutive relations are
\be
P/\rho = \Re T/\mu\quad ,\quad \kappa (\rho ,T) = \kappa_o(X_i)\rho^nT^{-s}\quad ,\quad
\varepsilon (\rho ,T) = \varepsilon_o(X)\rho^\lambda T^\nu\quad .
\ee
Expanded about the SSM, the exponents are: $n$ = 0.43, $s$ = 2.5, $\lambda$ = 1.0, $\nu$ = 4.2.
$\mu$, $\kappa_o$, and $\varepsilon_o$ are composition-dependent.  The luminosity constraint 
reads:~\cite{BP98,BK96}
\be
\f (pp) + (0.977)\f ({\rm Be}) + (0.751)\f ({\rm B}) + (0.956)\f ({\rm CNO}) =
6.55\times 10^{10}\ {\rm cm}^{-2}\ {\rm sec}^{-1}\quad .
\ee

The boundary conditions are imposed in a way appropriate to a {\em single} star: $M_\odot$, $L_\odot$, and 
$R_\odot$ fixed.  Homology then gives a family of possible non-convective {\em interiors} consistent with
observed outer solar features and parametrized by $\rho_c$ and $T_c$:
\bea 
\rho_c\sim\varepsilon^{-0.34}_o\kappa^{-0.40}_o\mu^{0.52}_cL^{0.085}_\odot\quad ,\quad
T_c\sim\varepsilon^{-0.13}_o\kappa^{-0.034}_o\mu^{0.22}_cL^{0.17}_\odot\quad .
\eea
The resulting $\f$ scale stably with $\rho_c$ and $T_c$ over a large range, 
giving the two-parameter homological mechanical/thermal variations of the SSM: 
$\f (i)\sim\rho^{\alpha_i}_c\cdot T^{\beta_i}_c$, with $(\alpha_i,\beta_i)$ for $pp$, Be, and
B $\nu$'s being $(-0.1,-0.7)$, $(0.7,9)$, and $(0.3,21)$, respectively.
The highest reactions in the $pp$ chain have the famous extreme sensitivity to $T_c$, while
all sensitivities to $\rho_c$ are mild and arise from the small luminosity contribution
made by the $ppII$, $ppIII$, and CNO chains.  It should be stressed that $\rho_c$ and $T_c$ are
model {\em outputs}, like the $\f (i)$.  These exponents reproduce the 1000-SSM Monte Carlo
and clarify that the {\em entire homological class} of SSMs has the wrong pattern of fluxes to explain
the {\em observed} energy dependence of $\phi:$ {\em lower} energies are {\em more} suppressed.
(Variation of nuclear cross sections also fail to explain the pattern.)
This conclusion depends only on $ppI$-dominance in $\varepsilon$, radiative diffusion in
$\kappa$, and the ideal gas law.

It is instructive to compare these homology results with the approximations often used to model
stars.  Perhaps the simplest (after the constant-density case) is the {\em Eddington standard
model}, based on a constant ratio of radiation to matter pressure throughout the star and equivalent
to an $(n,\gamma)$ = $(3,4/3)$ polytrope.~\cite{Arnett}  
This model, once its free parameters are fit, represents many main sequence stars not badly.
Homology applied to the mechanical structure alone automatically leads to a polytrope.~\cite{KW}
But the Bludman-Kennedy homology is more general than a polytrope, as it applies to both mechanical
and thermal structure.  It also scales
correctly in the evolved core, where the molecular weight $\mu$ changes
substantially, reflecting the chemical evolution that makes the present Sun differ from its
zero-age incarnation.  No polytrope fits this behavior.~\cite{BK96,BK99}

\begin{figure}[t]
\begin{center}
\epsfxsize=17pc 
\epsfbox{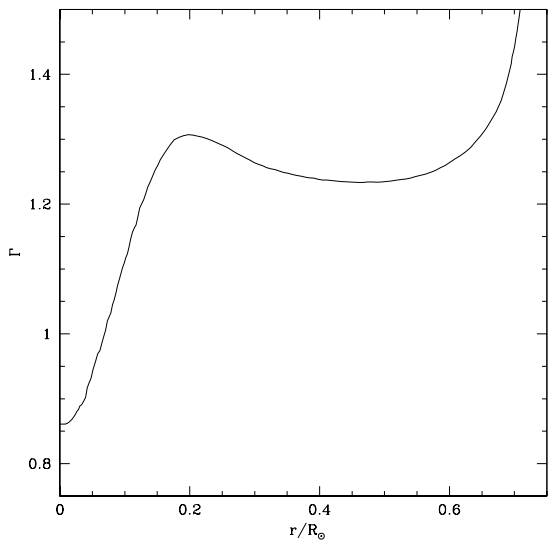} 
\end{center}
\caption{Stiffness profile $\Gamma (r)$ in the Bahcall-Pinsonneault 1998 SSM.~\protect\cite{BP98,BK99}
\label{fig:gamma}}
\end{figure}

A more complete version of homology is possible if the {\em differential} structure
is retained and rewritten using scale-invariant {\em homology variables}.~\cite{BK99}
For the entire mechanical, thermal, and chemical structural system,
the dimensionless differential equations are not less complex than
a full SSM.  But if the structure is restricted to the mechanical alone and a
barytrope $P(\rho )$ assumed, simple dimensionless structure equations follow.  
The key to the mechanical structure turns out to be the $\Gamma$ profile (Figure~\ref{fig:gamma}).
Constant $\Gamma$ gives a
polytrope again; in fact, $\Gamma_{\rm SSM}\simeq$ 4/3 outside the inner core, 
up to the CZ, where it rises to 5/3 (the adiabatic value).  But within the core, $\mu$ rises
and $\Gamma$ drops towards to the center, where $\Gamma^{\rm SSM}_c\simeq$ 8/9.
Described in terms of a polytrope, the effective index $n_{\rm eff}$ = $(\Gamma -1)^{-1}$ rises
in the core, diverging at $\Gamma$ = 1.  Further towards the center, $n_{\rm eff}$ rises from
minus infinity to a finite negative value at the center.  Such behavior in the inner core is not
even approximately polytropic and explains why attempts to use polytropes
to approximate the Sun only work (and only crudely) over the {\em whole} Sun and 
fail badly in the core.

\begin{figure}[t]
\begin{center}
\epsfxsize=17pc 
\epsfbox{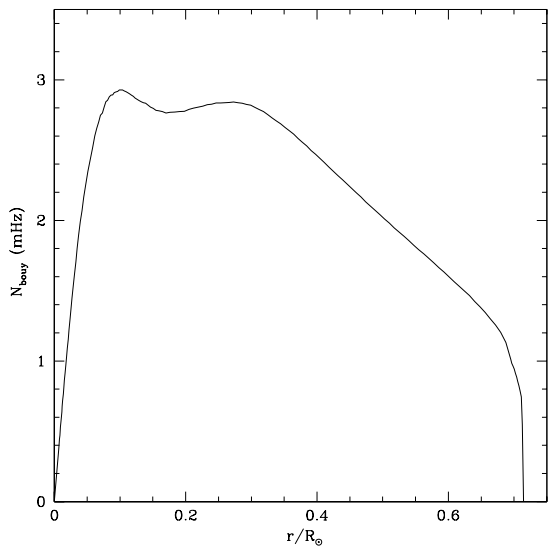} 
\end{center}
\caption{Bouyancy frequency profile $N(r)$ in the Bahcall-Pinsonneault 1998 SSM.~\protect\cite{BP98,BK99}
\label{fig:bouy}}
\end{figure}

The other approach to generalizing the SSM is to work with a few
reasonable assumptions, following these to simple, testable predictions.
Some powerful results are available, although restricted to mechanical structure only,~\cite{K00}
to which helioseismology is the key: adiabatic sound waves are mechanical perturbations with information
about sound speed, equation of state, and density and pressure profiles.~\cite{JCD97}  
Taking full advantage of these results
requires both $p$- (pressure) and $g$- (gravity) modes.  Their spectra can be inverted
to yield adiabatic sound speed $c_{\rm ad}(r)$ and bouyancy frequency $N(r)$ profiles, from
which follows the complete mechanical structure.  The BP98 $N(r)$ profile is shown in Figure~\ref{fig:bouy}.

The thermal and chemical structures cannot be directly probed by helioseismology, as their associated 
time scales are so long.  Simple homology implies $L\sim\mu^4M^3\kappa^{-1}_o
\varepsilon^{-0}_o$, but changes in $L_\odot$ large enough to explain
the solar $\nu$ deficit are probably ruled out by paleoclimatology.~\cite{paleo}
Direct tests of these aspects of the SSM require comparison to other Sun-like stars.
Such stars vary from the Sun somewhat in mass and chemical composition and
could lie anywhere on their respective evolutionary tracks.  Comparison properties
include luminosity, surface temperature, and photospheric radius.  With accurate 
photometry and parallaxes, accurate luminosities and colors are achievable.~\cite{iau111}  
The intermediary between these observations and stellar structure is stellar atmosphere models, which have
advanced considerably in the last 30 years.  Although still oversimplified, 
the models are good enough for solar-type stars to infer ranges for surface $T$, $g$, 
abundances, and turbulence.~\cite{Kurucz}  An exciting possibility will be opened by {\em asteroseismology} of 
Sun-like stars, as observation of stellar seismic modes (especially $g$-modes) would lead to direct 
characterization of stellar interiors.~\cite{BG94}

\section{Observational Issues}\label{sec:obs}

\begin{figure}[t]
\begin{center}
\epsfxsize=17pc 
\epsfbox{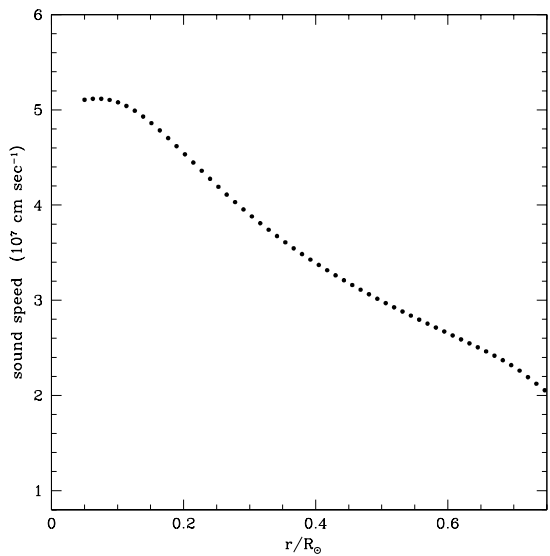} 
\end{center}
\caption{Solar sound speed profile $c_{\rm ad}(r)$ inferred from helioseismic observations.~\protect\cite{JCD97}
\label{fig:sound}}
\end{figure}

$M_\odot$, $L_\odot$, $R_\odot$, and surface $T$, as well as surface and proto-solar (meteoric) 
abundances, are well measured.  Helioseismic observations have directly or indirectly captured
millions of $p$-modes, allowing an accurate inversion of $c_{\rm ad}(r)$ down to $\x$ = 0.05 
(Figure~\ref{fig:sound}).~\cite{JCD97}
A convective core $(\gamma$ = 5/3) is ruled out, although circulation of heavy elements not
affecting heat transport cannot be at present.~\cite{CH}

The inferred sound speed peaks off-center at $\x\simeq$ 0.07.  Hydrostatic equilibrium requires $dc_{\rm ad}/dr$ = 0
at the center, but an off-center peak occurs where $\Gamma$ = 1.~\cite{BK99}  This peak and $dc_{\rm ad}/dr >$ 0 
for $\x <$ 0.07 indicate $\Gamma <$ 1 there,
a crucial confirmation of the core's chemically evolved state.
A complete profile down to the center becomes possible with the lowest $p$-modes.  But
complete inversion for model-independent mechanical structure would be possible only if the
higher $g$-modes were observed; analogous inversion would yield $N(r)$, and together $N$ and
$c_{\rm ad}$ yield $\Gamma$ and other mechanical profiles, the first truly independent
test of the SSM.~\cite{K00}

Comparing the Sun with other sun-like stars has been possible for many decades, albeit at poor 
precision.  But the recent Hipparcos-Tycho star catalogs (edition I in 1997: 1.1 M stars; 
edition II in 2000: 2.5 M stars) have revolutionized astrometry,
raising the accuracy of nearby stellar parallaxes by up to a factor of 10 or better (1 m-arcsec).~\cite{HippTyc}
Luminosities accurate to $<$ 5\% for nearby stars (within 25 pc) can now be inferred.  With good color
measurements and best current stellar atmosphere models, surface $T$'s can be limited to 1\%.~\cite{iau111,Kurucz}
Even more dramatic astrometric improvements could come from the proposed SIM and GAIA
orbital systems, to be launched in 2006 and 2009, respectively: 4 $\mu$-arcsec parallax errors
and luminosity errors limited only by photometry, tenths of a percent.~\cite{HippTyc}

\section*{Acknowledgments}

The author thanks the CSNP for the opportunity to present our results in honor of
Frank Avignone's life and work.  This work was done in collaboration with Sidney
Bludman (Univ. Pennsylvania and DESY) and Susana Tuzzo (Univ. Florida), who
helped sample and interpret the Hipparcos and SIMBAD catalogs, and
was supported by the U.S. DOE under grants DE-FG02-97ER41029 (Univ. Florida)
and DE-FG06-90ER40561 (Univ. Washington), the Institute for Fundamental
Theory (Univ. Florida), and the Eppley Foundation for Research.
The Institute for Nuclear Theory (Univ. Washington) and the Aspen Center for
Physics extended their hospitality.  The author is also grateful for important
interactions with John Bahcall (Institute for Advanced Study), J\o rgen
Christensen-Dalsgaard (Aarhus Univ.), Christopher Essex (Univ. Western Ontario),
Wick Haxton (Univ. Washington), Marc Pinsonneault (Ohio State Univ.), and Dimitri 
Pourbaix (ESA Hipparcos Science Team) concerning solar models, helioseismology, 
non-equilibrium thermodynamics, and the new classical astronomy of Hipparcos.


\begin{thebibliography}{99}

\bibitem{BHKL93}
S.~A.~Bludman, N.~Hata, D.~C.~Kennedy, and P.~G.~Langacker, \Journal{\PRD}{47}{2220}{1993};
N.~Hata and P.~Langacker, \Journal{\PRD}{56}{6107}{1997}.

\bibitem{BP98}
J.~N.~Bahcall, S.~Basu, and M.~H.~Pinsonneault, \Journal{\PLB}{443}{1}{1998};
M.~H.~Pinsonneault, private communication;
see also http://www.sns.ias.edu/\~{}jnb/.

\bibitem{TC98}
A.~S.~Brun, S.~Turck-Chi\` eze, and P.~Morel, \Journal{\ApJ}{506}{913}{1998}.

\bibitem{BK96}
S.~A.~Bludman and D.~C.~Kennedy, \Journal{\ApJ}{472}{412}{1996}.

\bibitem{BK99}
S.~A.~Bludman and D.~C.~Kennedy, \Journal{\ApJ}{525}{1024}{1999}.

\bibitem{K00}
D.~C.~Kennedy, {\em Astrophys. J.} {\bf 540} (2000), in press.

\bibitem{KW}
R.~Kippenhahn and A.~Weigert, {\em Stellar Structure and Evolution} (Springer-Verlag,
Berlin, 1990).

\bibitem{BU88}
J.~N.~Bahcall and R.~K.~Ulrich, \Journal{\RMP}{60}{297}{1988}.

\bibitem{Arnett}
D.~Arnett, {\em Supernovae and Nucleosynthesis: An Investigation of the History of
Matter, From the Big Bang to the Present} (Princeton Univ. Press, Princeton, 1996).

\bibitem{JCD97}
J.~Christensen-Dalsgaard, in {\em Proc. IAU Symp. No. 189}, ed. T.~R. Bedding, A.~J. Booth, 
and J.~Davis (Kluwer Academic, Dordrecht, 1997).

\bibitem{paleo}
T.~J.~Crowley and G.~R.~North, {\em Paleoclimatology} (Oxford Univ. Press, New York, 1991).

\bibitem{iau111}
{\em Proc. 111th IAU Symposium}, ed. D.~S.~Hayes, L.~E.~Pasinetti, and A.~G.~D.~Philip
(D. Reidel Publishing, Boston, 1985).

\bibitem{Kurucz}
See http://kurucz.harvard.edu/.

\bibitem{BG94}
T.~M.~Brown and R.~L.~Gilliland, \Journal{\ARAA}{32}{37}{1994}.

\bibitem{CH}
A.~Cumming and W.~C.~Haxton, \Journal{\PRL}{77}{4286}{1996}.

\bibitem{HippTyc}
M.~A.~C.~Perryman {\em et al}, {\em ESA Hipparcos and Tycho Catalogues SP-1200}, 17 vols.
(ESA Publications, Noordwijk, The Netherlands, 1997); see also
http://sci.esa.int/hipparcos/, http://astro.estec.esa.nl/GAIA/,
and http://sim.jpl.nasa.gov/.

\end{thebibliography}
\end{document}